\documentclass[sigconf]{acmart}
\usepackage{enumerate}
\usepackage{paralist}
\usepackage{multirow} 

\AtBeginDocument{%
  \providecommand\BibTeX{{%
    \normalfont B\kern-0.5em{\scshape i\kern-0.25em b}\kern-0.8em\TeX}}}

\acmYear{2023}

\iftrue
\newcommand{\jim}[1]{\todo[inline, color=green!40]{\textbf{Jim:} #1}}
\newcommand{\memo}[1]{\todo[inline, color=orange!40]{\textbf{Memo:} #1}}
\newcommand{\notes}[1]{\todo[inline, color=gray!40]{\textbf{Notes:} #1}}
\newcommand{\task}[1]{\todo[inline, color=orange!15]{\textbf{TODO:} #1}}
\newcommand{\assignedtask}[2]{\todo[inline, color=orange!15]{\textbf{TODO [#1]:} #2}}
\else
\newcommand{\jim}[1]{}
\newcommand{\notes}[1]{}
\newcommand{\memo}[1]{}
\newcommand{\task}[1]{}
\newcommand{\assignedtask}[2]{}
\fi

\begin{document}

\title{An Undergraduate Consortium for Addressing the Leaky Pipeline to Computing Research}
\author{James  Boerkoel}
\email{boerkoel@hmc.edu}
\affiliation{%
  \institution{Harvey Mudd College}
  \streetaddress{301 Platt Blvd.}
  \city{Claremont}
  \state{CA}
  \country{Country}
  \postcode{}
}

\author{Mehmet Ergezer}
\email{ergezerm@wit.edu>}
\affiliation{%
  \institution{Wentworth Institute of Technology}
  \streetaddress{550 Huntington Ave}
  \city{Boston}
  \state{MA}
  \country{Country}
  \postcode{}
}




\renewcommand{\shortauthors}{Omitted, et al.}


\begin{abstract}
Despite an increasing number of successful interventions designed to broaden participation in computing research, there is still significant attrition among historically marginalized groups in the computing research pipeline. 
This experience report describes a first-of-its-kind Undergraduate Consortium (UC) that addresses this challenge by empowering students with a culmination of their undergraduate research in a conference setting.
The UC, conducted at the AAAI Conference on Artificial Intelligence (AAAI), aims to broaden participation in the AI research community by recruiting students, particularly those from historically marginalized groups, supporting them with mentorship, advising, and networking as an accelerator toward graduate school, AI research, and their scientific identity.
This paper presents our program design, inspired by a rich set of evidence-based practices, and a preliminary evaluation of the first years that points to the UC achieving many of its desired outcomes.
We conclude by discussing insights to improve our program and expand to other computing communities.

\end{abstract}



\maketitle

\section{Introduction}
Equitable representation of all gender, racial, and ethnic identities is an ongoing challenge in computing despite annual increases in post-secondary enrollment \cite{camp2017generation}. 
While many interventions aimed at broadening undergrad participation in computing have achieved notable outcomes \cite{Giangrande2009,
Peckham2007,Russell2007,Alvarado2012,Alvarado2014}
, undergraduates often struggle to see a landing spot from which to launch their research careers amid the many systemic barriers they face.
This leads to significant attrition among students who identify as women; nonbinary; American Indian/Alaska Native; Black/African American; Native Hawaiian/Pacific Islander; Multiracial, not Hispanic; and/or Hispanic, any race (hereafter referred to as students from historically marginalized groups, HMGs\footnote{We acknowledge the term "historically marginalized group" imprecisely captures the nuances and intersections of identities that computing has and continues to marginalize \cite{williams2020underrepresented}.
We use this term to denote that marginalization is an active imposition by the dominant culture and to be consistent with the Taulbee survey and UC evaluations. 
However, discussing these identities in unison does not aim to equate the varied experiences or suggest one solution that meets the needs of individuals within those identities.
}).
In the United States and Canada during the 2019-2020 academic year, residents who identified as belonging to one or more HMG comprised 24.7\% of the Bachelor's degrees awarded in CS, but only 12.1\% of the Ph.D. enrollments and 9.6\% of the Ph.D.s awarded \cite{zwebenbizot2021}. 
These inequities persist in artificial intelligence (AI) \cite{zhang2021ai}.

The Undergraduate Consortium (UC)\footnote{\label{note1}The conference and event names are anonymized for review.} at the Association for the Advancement of Artificial Intelligence (AAAI)\footnotemark[2] aims to broaden participation in AI research by recruiting undergraduates, particularly those from HMGs, and supporting them with mentorship, advising, and networking as an accelerator toward graduate school, AI research, and their scientific identity.
Attracting and supporting undergraduates at highly-competitive technical conferences is both essential and challenging \cite{Davis2017}, mainly because such venues offer limited opportunities for undergraduates, leaving students questioning whether their identities and passions are relevant, valued, and if they belong. 
In this paper, we contextualize the broader landscape of evidenced-based practices for broadening participation in computing (BPC) that inform our program design.
Then we present our program design, evaluation, and results from the first years of the UC.
Finally, we report lessons learned and provide resources for implementing a UC at other STEAM conferences.


\section{Background} 
\label{sec:background}

Various compounding factors have contributed to disparities for students from HMGs over time, such as less access and encouragement to pursue education in computing \cite{2020state,gallup_2021}; social environments and structural issues of discrimination, bias, and othering within computing \cite{rankin2020intersectional,
cheryan2009ambient,mcgee2020interrogating}; and a lack of acceptance, understanding, and support of students' intersectional identities \cite{rankin2020intersectional,charleston2014intersectionality,kramarczukfirst, jaumotwomen}. 
These factors negatively impact students' self-efficacy, sense of belonging, computing identity, and science capital; constructs shown to predict the persistence of students from HMGs in computing research pathways \cite{nyame2015understanding,rorrer2018national,camp2020applying,lewis2017fitting}.

A breadth of strategies supports the
persistence in computing research for students from HMGs, such as research experiences for undergraduates (REUs) \cite{laursen2010undergraduate,peckham2007increasing,hathaway2002relationship}, 
attendance of technical conferences \cite{wright2019can}, and 
culturally-relevant research with real-world applications and societal impact \cite{rorrer2018national,isenegger2021understanding,vakil2018ethics}.
Mentorship from faculty and other senior members of the field, especially when it is inclusive of students' multiple identities \cite{spencerrole}, directly supports student interest and outcomes in research and graduate school. 
Research mentoring in computing is positively associated with the quality of graduate programs students enter \cite{cohoon2004mentoring}, overrides the difference in the sense of belonging between students from HMGs and the overrepresented majority (ORM) \cite{stout2018formal}, and strengthens graduate student self-efficacy \cite{tamer2016,tenenbaum2001mentoring}.
However, not only are students from HMGs less likely to have access to these mentors \cite{barker2009student,kim2011engaging}, underrepresentation of all genders, races, and ethnicities within senior members of the field necessitate cultural competence and intersectional mentoring capacity among all members of the computing research community to actively and consistently demonstrate allyship, regardless of mentor or student identities \cite{washington2020twice,spencerrole}.

Active engagement in a peer community with a breadth of intersectional identities helps students from HMGs grow self-efficacy and belonging by building social capital and validating their experiences  \cite{alvarado2019evaluating,jaumotwomen}.
This is particularly effective when that community represents a counterspace to the dominant cultures and structures of computing research that marginalize students who do not identify as Asian or white males \cite{ong2018counterspaces,stout2017grad,rorrer2021understanding}. 
Demonstrating and practicing technical, research, and professional skills in a safe and collectively defined environment is a crucial feature of these counterspaces being effective in computing research \cite{alvarado2019evaluating,rorrer2020ecsr}.

The UC fills an important gap in the current computing research pipeline.
The UC is well-situated to partner with the  growing opportunities for engaging undergraduates who identify as HMG in early research \cite{alvarado2019evaluating,laursen2010undergraduate,hathaway2002relationship}, 
celebrating undergraduate research, (e.g., the ACM Student Research Competition), or exposing undergraduates to career opportunities within a field (e.g., College Days at the RSA Conference). 
However, these tend to understandably sandbox undergrads from experiencing the institutional barriers that have historically excluded them, leaving a jarring transition to graduate research.
By partnering with AAAI, the UC enables students a scaffolded next step in their transition to graduate research by attending a  top, international AI research conference and giving broad exposure to and direct engagement with the latest ideas and top researchers in AI \emph{as a fellow research peer}.
The UC is the first known venue that offers specific programming to equip undergraduates to navigate the academic conference experience as a full member of the research community by cultivating a safe environment for practicing and refining skills and providing tools for a successful graduate research career.
The UC is also well-positioned to help future graduate students learn to leverage existing support systems.
A survey of eight top 2021 AI / AI-adjacent conferences 
demonstrates a clear trend toward offering opportunities to celebrate diversity and inclusion among student conference attendees (doctoral consortia, student activities programs, student research competitions).




\section{Program design}
\label{sec:program-design}
Our program design adopts and adapts the rich set of evidence-based BPC practices highlighted in Section \ref{sec:background}.
We designed the UC to achieve its goals by scaffolding the following components, each associated with a role within the UC leadership team.
The \emph{UC Chairs} act as traditional conference/event chairs, managing the program and ensuring that the various components come together effectively.
The \emph{Outreach Coordinator} builds awareness of the program (Subsection \ref{sec:outreach}).
\emph{Program Chairs} are responsible for launching the Call for Participation (Section \ref{sec:call}), managing the review process (Section \ref{sec:review}), and putting forth a recommended cohort of scholars.
The \emph{Mentorship Coordinator} scaffolds the mentoring programs describes in Subsection \ref{sec:mentor}.
The \emph{Student Engagement Coordinator} designs student support, activities, and logistics (Subsection \ref{sec:engagement}).
Finally, a \emph{Platform and Communications Coordinator} manages the various websites and tools for outreach, submission, and review processes.
As highlighted in Section \ref{sec:uciab}, curated templates of all program materials are publicly available.

\subsection{Targeted outreach}
\label{sec:outreach}
Acceptance decisions are not based on students' identities or demographics (e.g., ORM students are not desk rejected, downscored, etc.).
Ensuring a diverse cohort relies on targeted outreach to faculty (e.g., REU site leads), institutions (e.g., minority-serving institutions), venues (e.g., Tapia, Grace Hopper), and organizations (e.g., AccessComputing, Black in AI, CAHSI, IAAMCS, LatinX in AI, WIML) where students from HMGs are engaging in computing and/or research. We encourage faculty and others proximate to undergraduates to amplify the call and encourage students to apply, in addition to direct-to-student outreach, to encourage students with lower confidence and sense of belonging who are less likely to self-nominate for opportunities.
We repeat the UC's goal statement and target audience through all communications and provide email and social media templates for outreach.

Outreach occurs at three key intervals. 
Early notification is sent to faculty and professional organizations in May to raise awareness before the North American spring academic term ends.
We advertise the call for participation in July during the North American summer REU season.
A reminder is sent in September when students and faculty return at the top of the North American academic term, so that students can collaborate with advisors and on-campus resources to draft their submissions. 

\subsection{Call for participation}
\label{sec:call}
The Call for Participation (CfP) sets the program context and maps each application component directly to the review criteria.
The Personal Statement, Research Summary, and Advisor Feedback solicit qualitative information from students, enabling PC and UC chairs to identify the intended target audience.
Additional details that help support program logistics and evaluation are collected but not disclosed to PC Members during the review process. 

\subsubsection{Program Context}

The CfP opens by articulating its goal and reinforces it by describing: 
\begin{inparaenum}[(i)]
\item what the program provides students (mentorship, advising, networking, and travel support);
\item a list of anticipated activities; and
\item expectations of accepted students.
\end{inparaenum}
The CfP attempts to reassure students who may not see themselves as belonging to the AI research community by explicitly identifying the target audience, which includes undergraduates who: 
\begin{inparaenum}[(i)]
\item identify with HMGs in computing/AI/research; 
\item have contributed to an AI research project; and 
\item are at an inflection point where mentoring and networking will add unique value and support for their pursuit of AI research after graduation.
\end{inparaenum}
The CfP also highlights complementary opportunities at AAAI for students interested in forms of support other than mentorship (e.g., research challenge, scholarship, volunteering).
Finally, the CfP provides details about deadlines and event logistics.

\subsubsection{Personal Statement}

Applicants prepare a 2-3 page personal statement that discusses their AI research career journey thus far and highlights their individual contributions to an AI research project. 
We align discussion prompts to the review criteria to help students who lack experience writing personal statements (e.g., interest in AI research and career, societal impact, leadership skills, barriers overcome, role in the research project, goals for the UC). 
The personal statement format is inspired by the National Science Foundation Graduate Research Fellowship Program (NSF GRFP) so that students practice and receive constructive feedback on writing compelling personal statements for future applications. 

\subsubsection{Research Summary}

Applicants provide a two-page extended abstract that summarizes one of their significant AI research projects, including the research questions being investigated, highlighting crucial related work, the significance of their work, and possible future directions. 
We explicitly encourage students to seek their advisor's help reviewing and strengthening the summary.
The summary may describe collaborative research but must be written entirely by the applicant without unattributed passages written by others to evaluate the applicant's accomplishments rather than those of an advisor or research group.
Students followed the AAAI author kit and accepted students' research summaries are published in the AAAI proceedings, which enhances both the student's \textit{C.V.} and exposure to potential graduate advisors.

\subsubsection{Advisor Recommendation}
A graduate-degree-holding advisor validates the student's research contributions and speaks to how the applicant can contribute to and benefit from the UC. 

\subsection{Review process}
\label{sec:review}
The review of student submissions is managed by two volunteer Program Committee (PC) Chairs. 
Volunteer PC Members are solicited from faculty known to be committed to student development, mentorship, and broadening participation in AI. 
Each submission receives three reviews, and each PC Member reviews five submissions to ensure calibration in scoring. 
UC and PC Chairs conduct emergency reviews as needed.

The review criteria aim to evaluate student fit for the UC---students who are at a specific inflection point where mentorship will uniquely accelerate their research trajectory (i.e., completed some research, open to exploring post-graduation plans)---rather than overall intellectual merit, novelty, or contribution of the research presented as with typical conference submissions or undergraduate research competitions.
The review criteria map directly to the CfP to reduce subjectivity by ensuring PC Members use explicit parameters mapping to submission prompts.
PC members are instructed to consider biases that could impact their review and ensure that constructive feedback focuses on the statements rather than the individual.

Quantitative review feedback is confidential to the chairs and intended to assist in making decisions about whom to accept.
Quantitative feedback criteria are discretely defined in a rubric format on a scale from zero to four to increase alignment across reviewers and submissions.
The criteria are whether students demonstrate a potential to contribute positively to the UC cohort, whether they demonstrate an actionable readiness to leverage a UC mentor, whether the personal statement and research summary (effectively) includes all required components, and an overall final recommendation.
Scores from each of the three reviewers are averaged per criterion, summed, and stack ranked. 
Large standard deviations between reviewers are investigated to identify bias or poor calibration. 
UC and PC Chairs review the quantitative outcomes against the qualitative feedback provided by PC Members to identify the final student cohort.

PC members are requested to provide constructive, qualitative feedback to students on how they communicated and organized their statements by highlighting strengths and improvement opportunities.
UC and PC Chairs review qualitative feedback to ensure that it is constructive and unbiased; problematic feedback is stricken, and Chairs provide a new version.
The goal is to provide \emph{every} student with constructive research, writing, and career guidance from established professionals in the field. 

We solicit PC Member feedback after reviews conclude to understand whether the review process was straightforward, compelling, and well-supported; whether the number of submissions assigned, feedback requested, and time to review were sufficient; and whether interest in involvement in future UC cycles and roles.
This survey has resulted in immediate changes, such as streamlining the quantitative criteria and clarifying expectations in the CfP.

\subsection{Mentorship}
\label{sec:mentor}
The mentoring process is a fundamental part of the UC and provides undergraduates with not only support for their career development but also emotional support and role modeling \cite{jacobi1991mentoring,tenenbaum2001mentoring}. 
The UC leadership makes a concerted effort to ensure that our mentors and speakers represent the diversity we seek to achieve in our cohort and AI research.
We provide students with three levels of mentoring to target these mentoring functions.

\subsubsection{Faculty} 
We pair students with faculty members who are members of the AI research community and likely to be able to attend UC events.
This allows students to receive career advice and contacts that can help provide feedback as they pursue further graduate studies in AI from a leader in the field. 
Prospective mentors are invited to review applications to identify prospective mentees early and familiarize themselves with their research before matching.
Mentors and students are encouraged to meet before the event to discuss the student's goals and how to improve their submitted materials for publication, during the event to discuss career aspirations and general advice, and again after the event as needed to help support the student in the graduate school application process.
Mentors and students are provided with a broad set of discussion prompt to help guide discussion but are encouraged to tailor their conversations based on the student's needs.

\subsubsection{Near-peer}
The UC includes graduate student panelists and mentors who have recently succeeded in their doctoral program admissions, which will consist of UC alums in the future. 
These near-peer mentors provide socioemotional support, and their friendly counseling may appear more approachable to undergraduates who may struggle with feeling intimidated by more senior researchers.

\subsubsection{Peer}
Many of our networking activities and presentation feedback exercises provide a reciprocal peer-to-peer mentoring environment. 
These interpersonal relationships formed from the students' shared experiences generate a supportive cohort as they become members of the AI research community.



\subsection{Program of Events \& Student Engagement}
\label{sec:engagement}

The programming objectives of the UC are that every student who participates builds their identity as a scientist, expands their professional network, and learns networking and presentation skills. 
A COVID-19 pandemic-induced switch in modality from in-person to online caused the programming to evolve from the first UC in 2020 to subsequent years. 
All years share a common structure and events, such as two keynote speakers, a faculty mentor panel, a graduate student panel, and UC and AAAI poster sessions. 
The differences include a mentor/mentee offsite breakfast and lunch in 2020 and an online poster practice and other professional development exercises in 2021 and 2022. 

\subsubsection{Identity Building}

The keynote speakers invited to the UC addressed topics of interest for young researchers and provided academic empowerment among participants and tips for success in their graduate careers. 
These professional trajectories have served as examples for students to build their identities as researchers. 
Talk topics and themes have included overcoming imposter syndrome and stereotype threat, how to thrive on your path towards a Ph.D., and roles for computing in social justice.


\subsubsection{Community Building}

In 2020, the UC sponsored outings with academic faculty/mentors and social events with other student and affinity groups attending AAAI, whereas in 2021 and 2022, these connections happened virtually.
Graduate student panels across all years provided a faculty-free environment for undergraduate and graduate students to discuss what graduate school is ``really like''.
UC students across all program years are engaged in an online communication platform to provide peer-to-peer mentorship and support each other before and after the day of events. 
AAAI separately provided a robust student activities program, which we leveraged and encouraged students to attend, such as roommate pairing, receptions, and social programs. 
The AAAI 2020 student reception was held after the UC and allowed participants to network with other AAAI student attendees as a pre-formed cohort.

\subsubsection{Skill Building}

The personal statement and research summary, written by all applicants, receives constructive feedback to develop students' written communication skills. 
Students present their work and receive feedback from other undergraduate researchers, faculty, mentors, and the general AAAI audience. 
Participants and mentors discuss ways students could improve their presentations and statements during mentoring sessions.
Finally, the UC program of events included sessions designed to help students better recognize and articulate their strengths and passions as part of their statements.
There were also staged practice poster sessions that allowed students to iteratively improve their presentation based on feedback from a safe, supportive, and constructive audience. 

\section{Program Evaluation}
Evaluation of the UC is comprised of pre- and post-surveys for accepted students, delivered in the two weeks before and after the UC.
The student surveys measure students' self-efficacy, sense of belonging, computing identity, scientific capital, and professional skills, generally and relative to their experience at AAAI. 
The overall evaluation questions underpinning the student surveys are whether the UC accomplishes its stated goals and whether those accomplishments persist across demographic groups.
They also capture demographic information and students' qualitative feedback on the UC. 
The survey instruments and scales were developed for the UC. 
Changes from pre- to post-survey were measured using paired-sample t-tests and we use Cohen's $d$ to measure effect size. 

\subsection{Quantitative Results}
28 of the 39 students accepted to the UC program from 2020 to 2022, (72\%), had their pre- and post-survey responses matched.
19 of the 28 are from respondents who identify with one or more HMG (Table~\ref{tab:demographics}).
Additionally, we had students self-report physical disabilities, and six identified as first-generation and eight as international college students.

\begin{table}[]
\caption{Linked-sample student survey respondent demographics (2020, 2021, and 2022; n = 28).}
    \label{tab:demographics}
\resizebox{\columnwidth}{!}{%
\begin{tabular}{l|ccc}
    \toprule
Race/ethnicity                   & Man & Woman & Non-conforming \\
    \midrule
American Indian/Alaska Native    & 0   & 0     & 0                     \\
Asian or Asian American          & 5   & 6     & 1                     \\
Black/African American           & 2   & 2     & 0                     \\
Hispanic/Latinx                  & 0   & 0     & 0                     \\
Native Hawaiian/Pacific Islander & 0   & 0     & 0                     \\
White                            & 3   & 5     & 0                     \\
Two or more races/ethnicities    & 0   & 2     & 0                     \\
Not reported                     & 0   & 1     & 0                     \\
\midrule
First-gen & 3 & 3 & 0 \\
  \bottomrule
\end{tabular}
}
\end{table}

\begin{table}[]

\caption{Overview of pre/post survey items. Statistically significant differences in linked-sample student attitudes (2020-2022; n = 28) are bolded. Items that exhibited a large effect size (Cohen's $d>0.8$) across all three years are marked with *, items that dropped to medium effect size (Cohen's $d>0.5$) during the third year are marked with $\dagger$.
}
    \label{tab:significant}
\resizebox{\columnwidth}{!}{%
\begin{tabular}{l}
\toprule
                                                     
\textsc{To what extent are the following statements true of you:}\\
\textbf{I am comfortable navigating an academic conference}*                                
 \\
\textbf{I am confident interacting with other students at AAAI}*                                \\
\textbf{I am confident interacting with researchers at AAAI}$\dagger$                               \\
\textbf{I know what it is like to be an AI researcher}*  \\
\textbf{I know what career options are available to me in AI}*                                    \\
\textbf{I am comfortable talking about my research to faculty members} \\
\textbf{I have a strong sense of belonging to a community of researchers}* \\
I think of myself as a researcher  ($p=0.118$)     \\
\textbf{I am comfortable explaining the results of my research to faculty members}$\dagger$              \\
\textbf{I understand the resources available in AI to help me advance my career}*     \\
\textbf{I feel awkward in situations in which I am the only undergraduate student} \\

    \midrule
\textsc{How confident are you that you can:}\\
\textbf{Contact another researcher if I had a question about their area of expertise}* \\
Prepare a competitive application to graduate school ($p=0.083$)              \\
\textbf{Successfully articulate your research interests to a faculty member in the field}$\dagger$  \\
\textbf{Successfully complete a graduate school interview} \\                             
  \bottomrule
\end{tabular}}
\end{table}

Of 30 total responses to the post-survey, 20 (66.7\%) declared they had ``Very much'' achieved their self-determined goals for the UC, while the remaining 9 (30\%) felt they had ``Somewhat'' achieved their goals. 
Other strong results include consistently positive responses related to whether the UC helped build sense of belonging, actionable skills for advancing in computing research careers, and the ability to navigate AAAI confidently and productively.

Table~\ref{tab:significant} provides the language for the quantitative survey items for the pre-/post-event survey.
Outcomes were statistically significant (bolded) across 13 of 15 pre-to-post measures, every one of which exhibited at least a medium effect size (Cohen's $d>0.5$), and seven of which exhibited a large effect size (Cohen's $d>0.8$; noted with *). 
This points to the efficacy of the UC as an intervention across many of its goals.
Three additional items exhibited large effect sizes through the first two years of the program (marked with $\dagger$), but dropped to medium effect size when factoring in 2022, which generally saw smaller effect sizes across the board.
We attribute this slight drop to general pandemic and Zoom fatigue; indeed one student's suggestion for improving the event was to hold it in person.
2022 also abandoned using Slack as an asynchronous communication tool among the cohort, which led to student suggestions of clearer, more proactive communication and more opportunities for connecting with peers.
One of the impetuses behind our efforts to make resources available (Section \ref{sec:uciab}) is so lessons and effective tools persist through leadership transitions.

Results are not reported to prescribe the reliability of the student evaluation instruments for the broader CS education research community, but rather to understand the effectiveness of the UC as an intervention and the consistency with which respondents interpret and respond to the scales within. 
The intent of this experience report is to share the program model and learnings with early indicators from a small sample that it is accomplishing the intended goals.
Additional analyses and data are available upon request.

\subsection{Qualitative Student Feedback}

While the quantitative analyses provide a high-level view of the program efficacy, qualitative student feedback paints a more detailed picture of the impacts on personal research trajectories.
Responses highlight building sense of belonging:
\begin{quote}
    Dr. [omitted]'s talk on Overcoming Imposter Syndrome was truly impactful, and made for an inspiring start to the conference - following her talk, I felt like I belonged at AAAI and proceeded to put myself out there for the remainder of the conference.
\end{quote}
As well as practical advice from both panelists and mentors:
\begin{quote}
Getting insights on graduate applications by the AMA Panel was the best part of the workshop, apart from the invited talks. The advice given by the mentors was an irreplaceable part of the workshop!
\end{quote}
The presence of a supportive peer cohort proved vital:
\begin{quote}
UC social gatherings at conference events established through Slack were an excellent opportunity to make friends in an informal setting and to reflect on our common experiences as undergraduates at the conference - again, such conversations helped me feel like I belonged, and that I am capable of working alongside such fellow researchers as we collectively progress through graduate school and beyond. 
\end{quote}

Student comments pointed toward the overall impact and efficacy of the UC program. Various students remarks include:
\begin{quote}
    I have been admitted to multiple PhD programs and I attribute much of that success to the AAAI workshop and overall process. I will absolutely be pursuing my PhD in Fall 2020.
\end{quote}
\begin{quote} 
 I have a feeling that attending this conference \& consortium was one of the most impactful things I could have done for my future research career.
\end{quote}
\begin{quote}
     I think my most important takeaway from the UC was that I might have a future pursuing graduate education in AI/computer science and the importance of that experience in general.
\end{quote}
When students were asked for suggestions for improvements, besides the ``in person!!!!" requests made during the second virtual consortium, students asked for more interaction opportunities: 
\begin{quote}
Have more informal opportunities for UC scholars to interact with one another [...and with...] 
seniors who already applied
\end{quote}
And requested to make programming less dense:
\begin{quote}
The UC is great, but has a LOT of information to absorb. Maybe splitting it over two days instead of all in one day would be more beneficial.
\end{quote}
\section{Lessons and Insights}


\paragraph{Make Goals Explicit}
There are opportunities to make goals more explicit so that students can better appreciate and utilize them.
We evolved the CfP in this way, articulating who this is for, what is expected of participants, and what they will gain, as noted in Section \ref{sec:call}.
We punctuate this by also highlighting what the program is \emph{not}, and redirect students to more appropriate opportunities (e.g., AAAI student abstract or scholarship programs).
Anecdotal experience in the first two years showed that many students were hyper-focused on the quantitative review metrics. 
By streamlining the review process to only share qualitative, constructive, and actionable feedback with students and stating this goal more explicitly both in the CfP and review communications, we hope to mitigate the risk of students who are not accepted wondering ``why wasn't I good enough,'' and instead empower students with a better understanding of their strengths and growth opportunities as they continue their AI research journeys.

Similarly, we recognize that by being explicit about program goals, we recognize opportunities to target efforts to better achieve them.
An example is that our cohort is still overrepresented by students with strong support and connections within the AAAI community and could better target students from HMGs and students from community colleges and regional universities.
In the fall of 2022, we solicited demographic information of all applications in a way that was disassociated with their application. 
The small sample size for 2022 anecdotally points to the fact that our review process does a good job of retaining the diversity of our applicant pool when selecting the cohort, but that we can improve our recruitment of a more diverse applicant pool.
In the future, we aim to strengthen partnerships with existing programs for providing early research opportunities for students from HMGs, research competitions, and celebrations of diversity in computing to extend the reach of existing successful interventions\cite{csrmp,alvarado2019evaluating,rorrer2020ecsr, Boerkoel22}.

\paragraph{Metering Engagement}

Due to the scheduling and venue constraints of co-locating this event with a conference like AAAI, the entire program of events necessarily falls on a single day. 
While there might be limitations on extending formal programming across multiple days, 
we are exploring virtual/asynchronous programming ahead of the conference and holding the cohort-building activities at key intervals throughout the conference to allow students chances to discuss how to make the most of various aspects of their conference experience in a more ``just-in-time'' manner.
Similarly, while mentoring was designed to maintain a free-flowing, low-overhead relationship with well-renowned AI experts to humanize the idea of research, we recognize an opportunity to meter the intervals at which mentoring interactions happen to better complement the broader UC program through mentoring sessions with more well-defined topics and goals (e.g., early feedback on materials ahead of camera-ready deadlines, asking graduate school application questions closer to application deadlines, etc.). 

\paragraph{Stakeholder Education}
There are opportunities to better educate possible stakeholders about our program.
The goals of our review process deviate significantly from that of typical AAAI submission review processes, which led to a mismatch between our program's goals and some reviewers' expectations, particularly in the first year. 
We revised our CfP and review process to better spotlight the personal statement and provided much more explicit review instructions that attempt to orient reviews towards the goals of the UC.
Additionally, UC students' presentations are currently lost in a sea of other presentations.
There is an opportunity to more explicitly communicate and scaffold opportunities for potential graduate advisors to meet, interact with, and recruit UC scholars through mixer events or better highlighting presentations by ``prospective graduate students'' in the program.

\section{Sustainability and Portability}
\label{sec:uciab}
From its inception, the vision was to build the UC model to expand its reach by becoming a self-sustaining part of the AAAI conference and to replicate it at other computing and STEAM venues.
We have made significant strides in accomplishing both.
First, as detailed in Section \ref{sec:program-design}, we have taken stock of the various tasks and responsibilities associated with running the UC and divided them into a clearly articulated set of roles.
Second, we  documented the associated tasks and responsibilities with each role and provided templates for all communications and resources.
Third, we have provided a detailed schedule and mapping of all tasks across all roles into a detailed project plan that unfolds across the year.
These clearly-defined, well-scoped roles increase sustainability by expanding the community of individuals committed to the UC.

Further, we have packaged all of these UC materials in an accessible way so that other STEAM venues can easily adopt them.
We wrote all materials to be easily adaptable to other conferences by adjusting high-level deadlines to the rhythms of different venues and highlighting where discipline-specific knowledge might enhance the template materials.
The UC has been funded by a grant from the NSF along with support from [omitted for anonymity], and we include resources and advice for securing funding, including successful grant proposals.
The UC, grounded in evidenced-based strategies and constructs to broaden the participation of students from HMGs in computing and research, has produced promising preliminary results.
We hope to partner with others in the CS education community to help us expand the model to other technical conferences and define a more comprehensive evaluation infrastructure that validates the outcomes and their transferability.

\section{Discussion}


The UC design is inspired by a broad array of evidence-based practices from the CS education research community for broadening participation of students from HMGs in computing.
It aims to fill a need for supporting undergraduate researchers in their transition to graduate research by learning to navigate premier technical conferences. The UC provides a meaningful opportunity for students to present their undergraduate research, network with a community of peers, receive mentorship from senior researchers, and practice professional skills in a safe, alternative environment that builds their social and scientific capital.

The UC is an ongoing effort to strengthen the pipeline to computing research. This experience report highlights our undertaking to design, finance, and execute an event that \emph{any} STEAM venue could deploy. 
Students attending the UC showed significant gains with large effect sizes across key constructs known to support the pursuit and persistence of students from HMGs in computing research (self-efficacy, sense of belonging, computing identity, science capital).
These gains and encouraging student reflections strongly signal that the UC is accomplishing its goals.
The prevalence of students from HMGs in the UC student cohorts from 2020 to 2022 is an encouraging signal that the outreach and review strategies successfully attract and identify the target audience.
In 2022 we began to collect opt-in demographic data of all applicants to ensure that rates of representation across identities persist through to the accepted cohort.
Additionally, future cycles of the UC will collect opt-in data on students' disability status.
Our outreach strategy must stay responsive to include various communities and organizations as they continue to evolve in the computing and AI research fields.
Finally, we hope to establish new partnerships across other STEAM research communities and other programs that aim to broaden participation at other stages of the computing research pipeline to broaden and deepen the UC impacts.

\clearpage

\bibliographystyle{ACM-Reference-Format}
\bibliography{main}









\end{document}